# Illustrating the Michelson-Morley experiment

Bernhard Rothenstein[1], Stefan Popescu[2] and George J. Spix[3]

1) Politehnica University of Timisoara, Physics Department, Timisoara, Romania
2) Siemens AG, Erlangen, Germany
3) BSEE Illinois Institute of Technology, USA

**Abstract.** *Considering that the rays in the Michelson-Morley interferometer perform the radar detection of its mirrors, we use a relativistic diagram that displays, at a convenient scale, their location and the path of the rays. This approach convinces us that the rays that come from the two arms interfere with zero phase difference without using the usual ingredient, length contraction.*

## 1. Introduction

Telemetry[1,2,3] is associated with the detection of the space-time coordinates of distant events by receiving light signals that have left the point where the event took place (photographic detection) or by sending light signals towards the point where the event takes place and receiving it back after reflection (radar detection). In the case of the photographic detection of a luminous profile we work with a convergent bundle of light rays whereas in the case of the radar detection we work with a divergent bundle of light rays. The bundle ends at the observation point in the case of the photographic detection but starts from it in the case of the radar detection.

The plane electromagnetic wave is a mathematical construction in which the rays are parallel to each other and perpendicular to the wave front. A point like source of light emits a spherical electromagnetic wave. At a very large distance from the source we can consider that the wave it emits is a plane wave. Relativistic telemetry could be operative in the plane wave as well.

An approach to relativistic telemetry, which is free of paradoxes, works with the events generated by the involved light signals as detected from two inertial reference frames in relative motion. One of them is the rest frame of the profile we detect and respectively of the source of light involved in the telemetry. Performing the Lorentz-Einstein transformations on the space-time coordinates of the corresponding events, we can derive the equation that describes the detected profile in the reference frame relative to which it moves as a function of the relative velocity between the two frames and proper physical quantities (distance and angles) measured in the rest frame of the source. The reference frames involved are K(XOY) and K'(X'O'Y'). The corresponding axes of the two reference frames are parallel to each other, the OX(O'X') axes are common and K'(X'O'Y') moves with constant velocity $V = \beta c$ relative to K, in the positive direction of the common axes. At the origin of time in the two frames $t = t' = 0$ the origins of the two frames are located at the same point in space.

We present the detection of a profile using either a plane electromagnetic wave propagating parallel to the O'X' axis or a plane electromagnetic wave propagating parallel to the O'Y' axis. Figure 1 shows the first case.

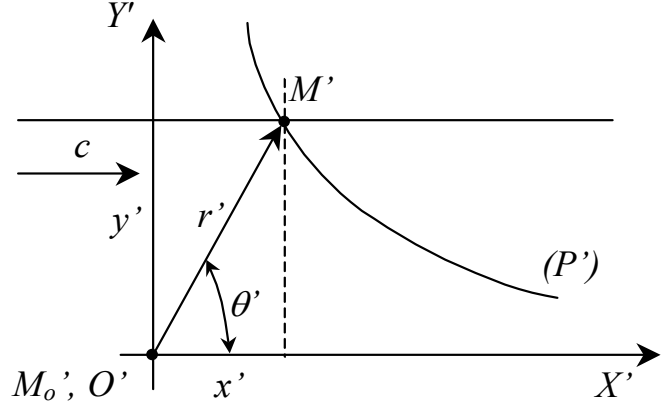

**Figure 1.** *A profile (P'), at rest in K', is located in a plane electromagnetic wave. The wave propagates in the positive direction of the O'X' axis.*

Arriving at the origin O' the ray that propagates along the common axes generates the event $E'_0(0,0,0)$. It is characterized in all inertial reference frames by the same space-time coordinates. A second ray of the same plane electromagnetic wave intersects the profile at the point $M'(x',y') = M'(r'\cos\theta', r'\sin\theta')$ and generates the event $E'(x',y',\frac{x'}{c}) = E'(r'\cos\theta', r'\sin\theta', \frac{r'\cos\theta'}{c})$, expressed using both Cartesian and polar space coordinates. The event $E'$ detected from K is characterized by the space-time coordinates $E(x,y,\frac{x}{c}) = E(r\cos\theta, r\sin\theta, \frac{r\cos\theta}{c})$. The Lorentz-Einstein transformations relate the space-time coordinates of the two events as:

$$x' = \gamma x(1-\beta) = r\sqrt{\frac{1-\beta}{1+\beta}}\cos\theta \qquad (1)$$

$$y' = r\sin\theta \qquad (2)$$

using the established relativistic notations $\beta = \frac{V}{c}; \gamma = (1-\beta^2)^{1/2}$. The polar coordinates of the corresponding events transform as:

$$r = r' \frac{\sqrt{1+\beta}}{\sqrt{1-\beta\cos 2\theta}} \qquad (3)$$

$$\tan\theta = \sqrt{\frac{1-\beta}{1+\beta}}\tan\theta' \qquad (4)$$

If the detected profile is in K' the circle

$$r' = R_0 \qquad (5)$$

then its shape in K will be described by:

$$r = R_0 \frac{\sqrt{1+\beta}}{\sqrt{1-\beta\cos 2\theta}} \,. \qquad (6)$$

We can transform the time coordinates of the events detected this way as:

$$t = t'\sqrt{\frac{1+\beta}{1-\beta}} \,. \qquad (7)$$

If the plane wave propagates in the negative direction of the common axes then we obtain the new equations for the transformation by changing the sign of c in the equations above. As we see all the equations are sensitive with respect to this change in the scenario.

The same profile is now located in a plane electromagnetic wave that propagates in the positive direction of the O'Y' axis, as shown in Figure 2.

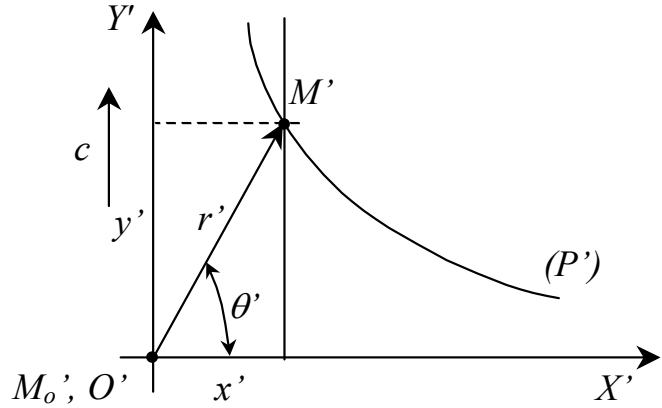

*Figure 2. A profile (P') at rest in K' is located in a plane electromagnetic wave. The wave propagates in the positive direction of the O'Y' axis.*

The ray that propagates along the O'Y' axis generates the event $E'_0(0,0,0)$ arriving at its origin that has the same space-time coordinates in all inertial reference frames. Another ray of the wave intersects the profile at a point $M'(x',y') = M'(r'\cos\theta', r'\sin\theta')$ generating the event $E'(x',y',\frac{y'}{c})$. Detected from K the same event is $E(x,y,\frac{y}{c})$. The Lorentz-Einstein transformations establish the following relationships between the space-time coordinates of the two events:

$$x' = \gamma x(1 - \beta \operatorname{tg}\theta) \tag{8}$$

$$y' = r\sin\theta \tag{9}$$

$$r = r'\frac{\sqrt{1-\beta^2}}{\sqrt{1-\beta\sin 2\theta}} \tag{10}$$

$$\tan\theta = \frac{\gamma^{-1}}{\dfrac{1}{\tan\theta'} + \beta} \tag{11}$$

We transform the time coordinates of the two events as:

$$t = \gamma t'\left(1 + \frac{\beta}{\tan\theta'}\right). \tag{12}$$

If the detected profile is the circle (5) then its shape detected from K will be described by:

$$r = R_0 \frac{\sqrt{1-\beta^2}}{\sqrt{1-\beta\sin 2\theta}}. \tag{13}$$

If the electromagnetic wave propagates in the negative direction of the O'Y' axis we obtain the corresponding results by changing the sign of $c$ in the equations derived above. All the equations are sensitive against this change in the scenario.

## 2. Relativistic diagrams

### 2.1. Relativistic diagram that displays in true magnitudes the results of relativistic telemetry with plane electromagnetic waves propagating parallel to the OX(O'X') axes.

The relativistic diagram we propose presents perpendicular axes on which we measure the space coordinates of the events involved as we show in Figure 3. It displays the circle $r' = R_0$ ($R_0 = 1$) and the curve described by (6) considering the case when the wave propagates in the positive direction of the OX(O'X') axes.

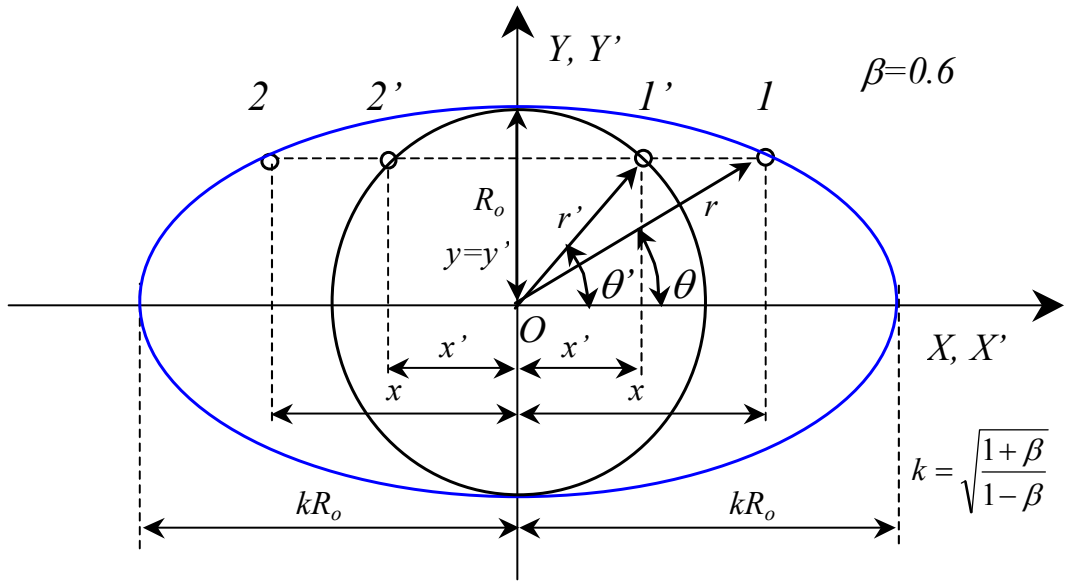

***Figure 3.*** *The relativistic diagram that enables us to detect from the reference frame K the shape of a circle $r' = R_0 (R_0 = 1)$ at rest in K', using the rays of a plane electromagnetic wave propagating in the positive direction of the common axes.*

The invariance of distances measured perpendicular to the direction of relative motion enables us to find out the location on the diagram of events $E$ and $E'$ as well as to measure on it the corresponding space coordinates in true magnitudes. Figure 4 shows the relativistic diagram valid for the case when the wave propagates in the negative direction of the OX(O'X') axes.

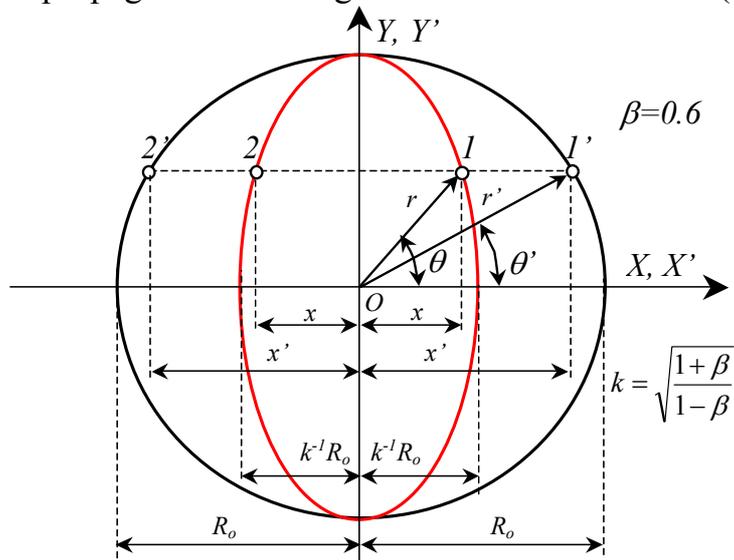

***Figure 4.*** *The relativistic diagram that enables us to detect from the reference frame K the shape of the circle $r' = R_0$ ($R_0 = 1$) at rest in K' using the rays of a plane electromagnetic wave propagating in the negative direction of the common axes.*

**2.2. Relativistic diagram that displays in true magnitudes the results of relativistic telemetry with plane electromagnetic waves propagating parallel to the OY(O'Y') axes.**

Figure 5 shows the relativistic diagram valid for the case when relativistic telemetry is performed using a plane electromagnetic wave propagating in the positive direction of the O'Y' axis. It presents perpendicular axes on which we measure the space coordinates of the events involved. It displays the circle $r' = R_0$ and its shape detected from K described by (13). The invariance of distances measured perpendicular to the direction of relative motion enables us to find on the diagram the location of events $E$ and $E'$ and to measure on its axes the corresponding space coordinates.

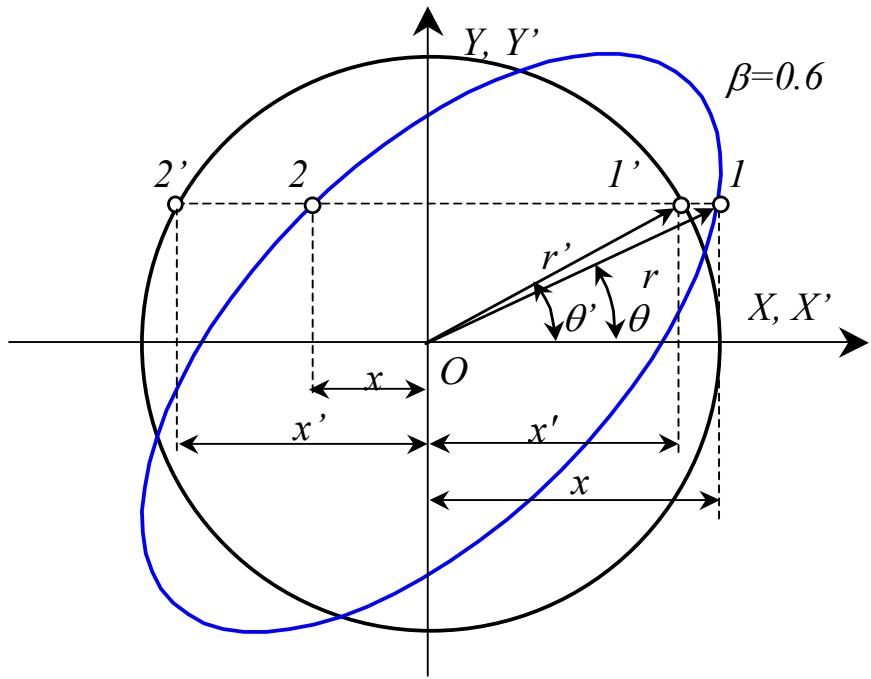

*Figure 5. The relativistic diagram that enables us to detect from the reference frame K the shape of the circle $r' = R_0$ ($R_0 = 1$) at rest in K' using the rays of a plane electromagnetic wave propagating in the positive direction of the O'Y' axis*

Figure 6 shows the situation when the wave changes its propagation direction.

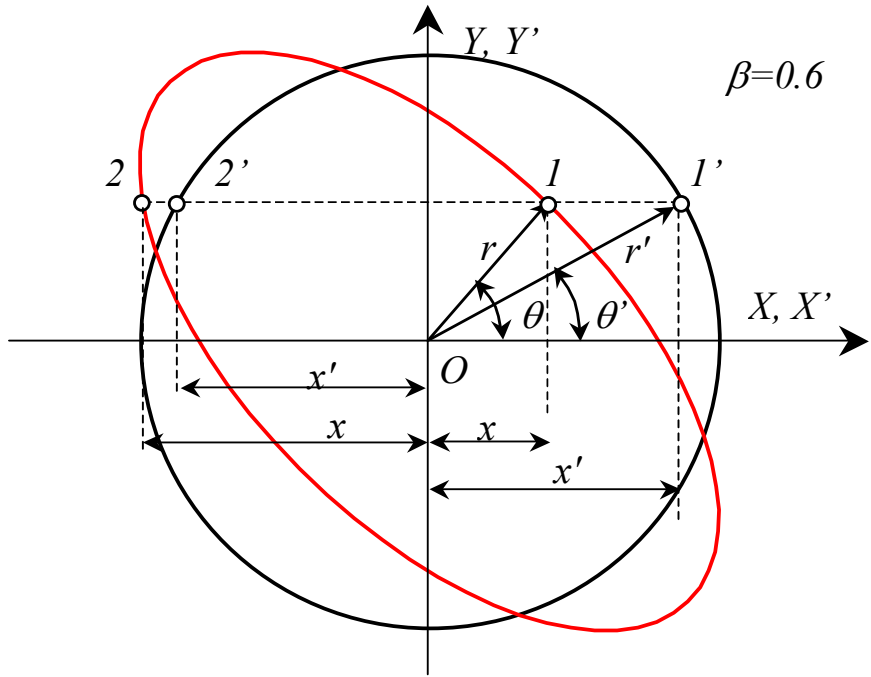

*Figure 6. The relativistic diagram that enables us to detect from the frame K the shape of the circle $r' = R_0$ ($R_0 = 1$) using the rays of a plane electromagnetic wave propagating in the negative direction of the O'Y' axis.*

### 3. Illustrating the Michelson-Morley experiment.

The Michelson-Morley experiment[4] is a strong convincing argument in the favour of Einstein's second postulate that is otherwise somehow counterintuitive for those who just start learning special relativity. Its importance in teaching special relativity as an introductory subject is largely debated in the literature[5].

We agree with Schumacher[6] by considering that the Michelson-Morley experiment should be taught to students who have the skill to handle the Lorentz-Einstein transformations. Our purpose is to illustrate how the Michelson-Morley interferometer works when we observe it from the reference frame K and we know how it works in its rest frame K'. We intend to illustrate this experiment using the relativistic diagrams we have presented so far. These diagrams convert the Lorentz-Einstein transformations into a movie and display in true magnitudes the Cartesian and polar coordinates of the events involved as well as the distances travelled by the light signals inside the interferometer arms. They should convince us that:

- The semi-transparent mirror at the centre of the interferometer correctly reflects the light to and from the interferometer axes and correctly overlaps the longitudinal ray over the transverse ray at the detector.
- The distances travelled by the light waves interfering at the observation point are equal to each other, the waves arriving there without phase difference.

Figures 7a and 7b shows the Michelson-Morley interferometer in action. Figure 7a depicts the situation when the half silvered mirror (HSM)' reflects the incident plane wave propagating in the positive direction of the common axes (ray *a'*) towards the mirror 1 $(M_1)$' located at the end of the vertical interferometer arm and permits it to propagate farther towards mirror $(M_2)$' located ate the end of the horizontal arm (ray d'). The origin of time *(t=t'=0)* coincides with the ray arriving time at the centre of the half silvered mirror.

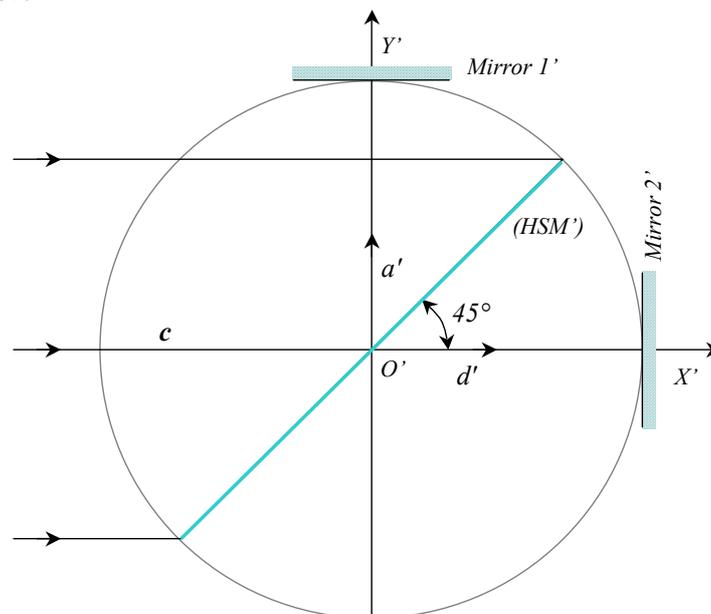

***Figure 7a.*** *The Michelson-Morley interferometer and its rays at the time when the ray incident at its centre is partially reflected towards mirror 1' located at the end of the vertical arm (ray a') and partially continues to propagate towards mirror 2'.*

Figure 7b depicts the situation when the ray *b'* reflected by the mirror $(M_1)$' and the ray reflected by the mirror $(M_2)$' located at the end of the horizontal interferometer arm arrive at the centre of the half silvered mirror. Without losing in generality we can make a time-shift considering that the arrival of the two rays at the centre of (HSM)' takes place at *t=t'=0*.

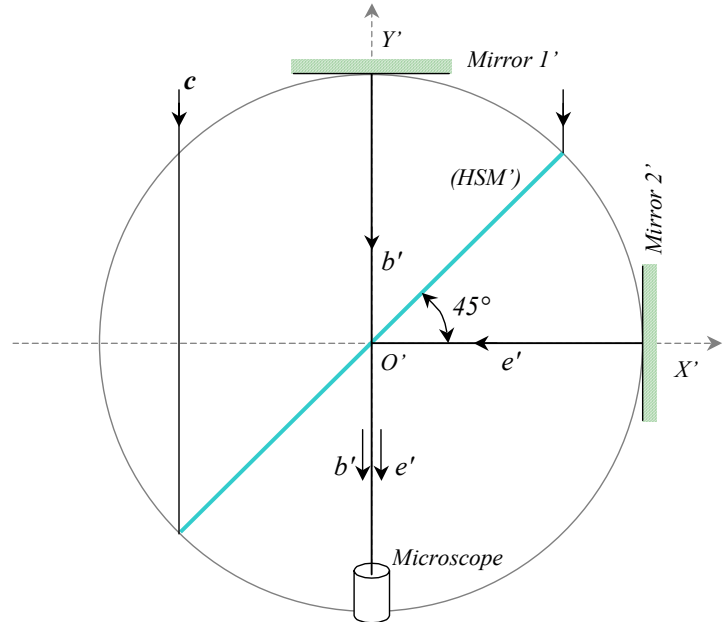

***Figure 7b.*** *The Michelson-Morley interferometer at the time when the ray b' reflected by mirror 1' and ray e' reflected by mirror 2' arrive simultaneously at the half silvered mirror. The rays b' and e' are directed towards a microscope where they interfere with a zero shift in time.*

In the situation depicted in Figure 7a we consider parallel rays of an electromagnetic wave propagating in the positive direction of the O'X' axis and detected at the half silvered mirror respectively parallel rays of an electromagnetic wave propagating in the positive direction of the O'Y' axis detected at the half silvered mirror and at the mirror 1' as well. In order to find out, on our relativistic diagram, the location of the events involved we present overlapped in Figure 8: the circle $r' = R_0$ ($R_0 = 1$), its shape detected from K using parallel rays propagating in the positive direction of the O'X' axis (6) and its shape detected from K while using light signals that propagate in the positive direction of the O'Y' axis (10). The rules of handling the diagram enable us to find out that if the segment 1'2' represents the half silvered mirror as detected from K' ($\Psi_{HSM'} = 45^0$) then the segment 12 represents its shape as detected from K' (HSM) making an angle $\Psi_{HSM}$ with the OX axis given by:

$$\tan \Psi_{HSM} = \sqrt{\frac{1-\beta}{1+\beta}}. \qquad (14)$$

$M_1$ represents the location of mirror 1' when detected from K.

The ray *a'* propagates in the vertical interferometer arm towards the mirror $(M_1)'$ located at its end. When detected from K this ray appears propagating along a direction $\Psi_a$ given by ($\theta' = 90^0$):

$$\tan \Psi_a = \frac{\gamma^{-1}}{\beta} \qquad (15)$$

The ray *a* propagating in the vertical arm travels over a distance:

$$OM_1 = \frac{L_0}{\sin \Psi_a} = \frac{L_0}{\sqrt{1-\beta^2}} \qquad (16)$$

whereas the ray that propagates through the horizontal interferometer arm travels over a distance:

$$OM_2 = L_0 \sqrt{\frac{1+\beta}{1-\beta}} \qquad (17)$$

$L_o$ representing the proper lengths of the arms.

Figure 9 shows the situation presented in Figure 7b as detected from K. In accordance with (4) the HSM makes an angle $\Psi_{HSM}$ with the OX axis given by:

$$\tan \Psi_{HSM} = \sqrt{\frac{1+\beta}{1-\beta}}. \qquad (18)$$

The ray reflected by $M_2$ and by the HSM (*e*) should be detected as propagating along the same direction as the ray *b,* because in K' they propagate along the negative direction of the O'Y' axis. Both rays propagate along a direction $\Psi_{b,e}$ given by:

$$\tan \Psi_{b,e} = -\frac{\gamma^{-1}}{\beta} \qquad (19)$$

and interfere in the microscope. The ray that propagates in the vertical interferometer arm travels over a total distance:

$$OM_1 + M_1O = \frac{2L_0}{\sqrt{1-\beta^2}}. \qquad (20)$$

The ray that propagates in the horizontal interferometer arm travels over a total distance:

$$OM_2 + M_2O = 2\frac{L_0}{\sqrt{1-\beta^2}}. \qquad (21)$$

As expected, the two rays travel over the same total distances arriving at the microscope without phase difference. This explains the negative result of the experiment. Simple algebra shows that the normal to the HSM is in the case presented in Figure 8 the bisector of the angle made by the rays *a* and ray *b*.

*Figure 8.* *This relativistic diagram enables us to detect the location of the mirrors and the path of the rays as detected from K in the situation depicted in Figure 7a. The segment 1'2' represents the location of the half silvered mirror as detected from K', while the segment 12 represents its location as detected from K. $M_1$ represents the location of mirror 1' and **a** represents the path of ray a' when detected from K. It displays the circle $r' = R_0$ ($R_0 = 1$), its shape in K as detected using parallel rays propagating in the positive direction of the common axes and its shape in K as detected by parallel rays propagating in the positive direction of the O'Y' axis. When detected from K the mirror 2' is located at $M_2$.*

Consequently in the case of Figure 9 the normal to the HSM is the bisector of the angles made by the rays *b* and *e*.

*Figure 9. This relativistic diagram enables us to detect the position of the mirrors and the path of the rays as detected from K in the situation presented in Figure 7b. $M_1$ represents the location of $(M_1)'$ and $M_2$ represents the location of $(M_2)'$. The ray b represents the path of ray b' while e represents the path of the ray reflected by mirror 2'. The segment 1'2' represents the half silvered mirror in K' and the segment 12 represents it as detected from K. The diagram displays the circle $r' = R_0$ ($R_0 = 1$) and its shape as detected using parallel rays that propagate in the negative direction of the common axes respectively with rays that propagate in the negative direction of the O'Y' axis.*

In the case depicted in Figure 8 the normal to the mirror makes with the OX axis an angle $\Psi_n$ given by:

$$\tan \Psi_n = -\frac{1}{\tan \Psi_{HSM}} = -\sqrt{\frac{1-\beta}{1+\beta}}. \tag{22}$$

Taking into account that:

$$\tan \theta = \frac{-1+\sqrt{1+\tan^2 2\theta}}{\tan 2\theta} \tag{23}$$

the bisector of the angle $\Psi_a$ makes with the OX axis an angle $\Psi_{bis}$ ($\Psi_{bis} = \dfrac{\Psi_a}{2}$) given by:

$$\tan \Psi_{bis} = -\sqrt{\dfrac{1-\beta}{1+\beta}}. \qquad (24)$$

Comparing (22) with (24) we see that $\Psi_n = \Psi_{bis}$ and therefore the reflection law on a plane mirror works in K and in K' as well. A similar analysis convinces us that the reflection law is valid in the situation depicted in Figure 9 as well.

### 4. Conclusions

Reducing the Michelson-Morley to radar and to a photographic detection of the half silvered mirror we propose a relativistic diagram that envisages the paths of the light rays in the interferometer as detected from the stationary reference frame. The diagrams display in true magnitudes the angles and the apparent shapes of the mirrors. Our analysis does not take into account the length contraction of the horizontal interferometer arm in contrast with Lorentz (1895) who considered that the negative result of the Michelson-Morley experiment could be explained by the contraction of the length of the horizontal arm. A qualitative diagram tracing the path of the rays is presented by Soni[7].